\documentstyle[twoside,fleqn,espcrc2]{article}

\input epsf

\title{Spin Chains and Chiral Lattice Fermions}
\author{
	H. B. Thacker\\
        Dept.~of Physics,
	Univ.~of Virginia, Charlottesville, VA~~22901~~USA
\thanks{UVA-INPP-94-5;hep-lat/9412019}}
\begin{document}
\begin{abstract}
The generalization of Lorentz invariance to solvable two-dimensional
lattice fermion models has been formulated in terms of Baxter's corner
transfer matrix. In these models, the lattice Hamiltonian and boost operator
are given by fermionized nearest-neighbor Heisenberg spin chain operators.
The transformation properties of the local lattice fermion operators under a
boost provide a natural and precise way of generalizing the chiral
structure of a continuum Dirac field to the lattice.  The resulting
formulation differs from both the Wilson and staggered (Kogut-Susskind)
prescriptions. In particular, an axial $Q_5$ rotation is sitewise local,
while the vector charge rotation mixes nearest neighbors on even and odd
sublattices.
\end{abstract}
% typeset front matter (including abstract)
\maketitle

%%%%%%%%%%%%%%%%%%%%%%%%%%%%%%%%%%%%%%%%%%%%%%%%%%%%%%%%%%%%%%%%%%%%%%%%%%
One of the most interesting developments in the theory of solvable
two-dimensional lattice statistics models is Baxter's invention of the
``corner transfer matrix'' (CTM)[\cite{baxtersbook}].
In the same sense that an ordinary
(row-to-row) transfer matrix can be viewed as an operator which translates
by one row of sites on a lattice, the CTM may be defined as
an operator which rotates around a central site from a horizontal row (or
half-row) of sites to the corresponding vertical row, or vice-versa. Thus,
for example, the partition function may be written as the product of four
CTM's. From this point of view, it is not
surprising that the CTM is associated with a lattice version of Euclidean
rotational or Lorentz symmetry. What is surprising is that, in certain
integrable models like the 8-Vertex/XYZ-spin chain, the CTM contains
a continuously variable anisotropy (``spectral'') parameter which plays the
role of rapidity or Euclidean rotation angle. The action of the full
one-parameter set of CTM's on the Hamiltonian eigenstates is
completely analogous to the action of the Lorentz boost operator in a
relativistically invariant continuum theory, i.e. it uniformly shifts the
rapidity of all the particles (spin-waves)
in the state by a continuously variable boost
velocity.  The implications of the CTM for the space-time symmetry
properties of integrable lattice models has been discussed elsewhere
\cite{HBT_CTM,Itoyama}. Here I want to show how the CTM formalism may be used
to
investigate the chiral structure of the lattice fermions obtained
from the 8-vertex/XYZ chain. It was shown long ago by
Luther (\cite{Luther}) that, in the scaling limit,
this model reduces to the relativistically invariant massive Thirring model,
a theory of self-coupled, massive Dirac fermions. Since the chiral structure
of the massive Thirring model may be expressed in terms of the Lorentz
transformation properties of the Dirac field, the CTM formalism provides a
natural and unique generalization of this chiral structure
to the lattice. To summarize the main result of this investigation, we find
that the XYZ spin chain Hamiltonian constitutes a formulation of lattice
Dirac fermions which differs essentially from both
the Wilson and Kogut-Susskind (KS) prescriptions. In particular, for both
Wilson and KS formalisms, the complex spinor
components of the Dirac field remain sitewise
local. (In the KS case, different spinor components are spread
over neighboring sites, but each site
supports complex fermion components.) Thus,
the conserved U(1) vector charge $Q = \int \bar{\psi}\gamma_0\psi dx$ of the
Dirac field simply generates a phase rotation at each site. By contrast, the
lattice prescription that we infer from the CTM analysis of the
vertex/spin-chain model leads to a local axial-vector charge but a
nonlocal vector charge operator which
mixes even and odd nearest neighbor fermion fields.
In fact, the conserved vector charge is not built into the
lattice spin-chain theory a priori (as in the Wilson and KS formalisms),
but only emerges as a consequence of fermion doubling. The nonlocality
of the vector charge results from
the reduction of the Brillouin zone, as I will outline here. A more
extensive discussion of vertex models, spin chains, and chiral lattice
fermions will be presented elsewhere.

The general, anisotropic nearest-neighbor Heisenberg chain Hamiltonian
may be written $H=\sum H_j$, where
\begin{equation}
H_j = -\frac{1}{2}\left\{\sigma_j^x \sigma_{j+1}^x +
k\sigma_j^y \sigma_{j+1}^y
+ \Delta \sigma_j^z \sigma_{j+1}^z\right\}
\end{equation}
The boost operator that emerges from the CTM analysis is the first moment of
the same density, $K = \sum[jH_j]$.
This model can be converted to a lattice fermion theory by a Jordan-Wigner
transformation
\begin{equation}
c_j^{x,y} =
\sigma_j^{x,y}\left(\prod_{j^{\prime}<j}\sigma^z_{j^{\prime}}\right)
\end{equation}
Here, $c^x$ and $c^y$ are real canonical fermion fields,
\begin{equation}
\left(c_j^{x,y}\right)^* = c_j^{x,y},\,\,\,
\left\{c_j^a,c_{j^{\prime}}^b\right\} = \delta_{ab} \delta_{jj^{\prime}}
\end{equation}
Expressing the Hamiltonian in terms of a single complex fermion at each site,
$c_j = \left(c_j^x+ic_j^y\right)/\sqrt{2}$,
the terms in $H$ can be written as a kinetic energy term, a mass term,
and an interaction term, respectively,
$H_j = T_j + M_j + V_j$, where
\begin{eqnarray}
T_j & = & \frac{1}{2}(1+k)\left(c_j^*c_{j+1} + c_{j+1}^*c_j\right) \\
M_j & = &\frac{1}{2}(1-k)\left(c_j^*c_{j+1}^* + c_{j+1}c_j\right) \\
V_j & = & -\frac{1}{2}\Delta c_j^*c_jc_{j+1}^*c_{j+1}
\end{eqnarray}
Note that the mass term violates the symmetry under a global phase
rotation of $c_j$ at each site. In the continuum limit, this global phase
symmetry is associated with the conservation of axial-vector current,
$J_5^{\mu} = \bar{\psi}\gamma_5\gamma^{\mu}\psi$.
{}From here on, let us focus on the free-fermion case $\Delta=0$.
For this case, the
eigenstates of $H=T+M$ may be written simply in terms of momentum-space fermion
operators,
\begin{equation}
a_{x,y}(z) = \sum_jz^j c^{x,y}_j
\end{equation}
The eigenmode operators of $H$ are given by
\begin{equation}
B(z) = z^{-1}C(z)a_x(z)+iC(z^{-1})a_y(z)
\end{equation}
where $C(z)=(1+kz^2)^{1/2}$. They satisfy $[H,B(z)]=\omega(z)B(z)$,
where the single particle energy is
$\omega(z) = C(z)C(z^{-1})$. The boost operator, defined as
$K = \sum j[T_j+M_j]$ has the fundamental property \cite{HBT_CTM,Itoyama,Sogo}
of inducing a uniform
rapidity shift on the Hamiltonian eigenmodes, as expressed by the
commutator
\begin{equation}
[K,B(z)] = z\omega(z)\frac{\partial}{\partial z}B(z).
\end{equation}

In order to understand the chiral structure of the local lattice fields,
we need to look at their behavior under a boost. When expressed in
momentum space, the relationship between the transformation
of the local fields and that
of the eigenmodes discussed above is quite analogous to the continuum
case. After some analysis, we find that the transformation of the local
field can be expressed in reduced form by defining
\begin{equation}
\psi_{1,2}(z) = \left(a_x(z)\pm i\sqrt{k}a_y(z)\right)/\sqrt{2}
\end{equation}
which satisfy
\begin{equation}
\label{eq:localboost}
[K,\psi_{1,2}(z)] = z\frac{\partial}{\partial z}[H,\psi_{1,2}(z)]
\pm z^{-1}\sqrt{k}\psi_{1,2}(z)
\end{equation}
This commutator provides a precise generalization of Lorentz transformations
to the lattice.
By inspecting the energy function $\omega(z)$ we see that there are two scaling
regimes as $k\rightarrow 1$, one at $z\approx i$ and one at $z\approx -i$.
In the first region, Eq. (\ref{eq:localboost}) reduces to
precisely the Lorentz transformation
of the chiral components of a Dirac field. Thus if we
restrict $z$ to the upper half
of the complex plane,  the lattice fields $\psi_{1,2}(z)$ are the chiral
components of the lattice Dirac field (in momentum space).
For the second scaling region at $z\approx -i$
(the ``doubler'' modes) the Lorentz structure is
inverted from what it should be if we
try to naively extend this identification to the LHP (e.g. by
taking $z\rightarrow -z$ the last
term in (\ref{eq:localboost}) changes sign). The proper way to deal
with the doubler modes is to define a reduced Brillouin
zone $0\leq arg\,z < \pi$ by using the reflection
relation that follows from the reality of the
$a_x $ and $a_y$ fermions, namely
\begin{equation}
\psi_{1,2}(z) = \psi_{2,1}^*(z^*)
\end{equation}
(In the continuum Dirac theory, this represents CP conjugation.)
Any operator involving fermion operators in the lower half z-plane
should be rewritten in
terms of those in the UHP (using the above reflection formula)
in order to correctly
exhibit its chiral structure. The doubler modes in the LHP, reflected into
the reduced Brillouin zone, provide the CP conjugate states relative to
those in the UHP. This has interesting implications for
the structure of the vector charge
on the lattice. For example, if we define the Fourier transform of the
local complex fermion,
$\psi(z) = \sum z^jc_j $, we must write this local fermion in terms of
Dirac fields of
indefinite vector charge. In the reduced Brillouin zone
$z\,\epsilon\,$UHP, we can write
the fermion fields on even and odd sites separately in terms of the chiral
fields (taking $k\rightarrow 1$),
\begin{equation}
\label{eq:evenodd}
\psi_{e,o}(z) \equiv \psi(z) \pm \psi(-z) = \psi_1(z) \pm \psi_2^*(-z^*)
\end{equation}
It may be verified directly that the Hamiltonian reduces to
that of a Dirac fermion with
chiral components as indicated.
(In the continuum limit, this calculation reduces to the
original analysis of Luther\cite{Luther}.)
\begin{eqnarray}
T & = & \frac{1}{2}(1+k)\int_{UHC}\frac{dz}{i\pi z}(z+z^{-1})\\& &\left[
\psi_1^*(z)\psi_1(z) - \psi_2^*(z)\psi_2(z)\right]\\
M & = & \frac{1}{2}(1-k)\int_{UHC}\frac{dz}{i\pi z}(z-z^{-1})\\& &\left[
\psi_1^*(z)\psi_2(z) - \psi_2^*(z)\psi_1(z)\right]
\end{eqnarray}
where the integrals are over the upper half of the unit circle.
Note in particular that with this reduction of the Brillouin
zone, the Hamiltonian
(including the mass term) exhibits a phase invariance under a
vector charge rotation,
$\psi_{1,2}\rightarrow e^{i\phi}\psi_{1,2}$.
This rotation mixes the local even- and odd-site fermions,
\begin{eqnarray}
\psi_e & \rightarrow & \cos\phi\psi_e+i\sin\phi\psi_o\\
\psi_o & \rightarrow & \cos\phi\psi_o+i\sin\phi\psi_e
\end{eqnarray}
Note however that under an axial-vector rotation,
$\psi_1\rightarrow e^{i\phi}\psi_1,\,\,\,\psi_2\rightarrow e^{-i\phi}\psi_2$,
the site fermions carry a definite chirality: $\psi_{e,o}\rightarrow
e^{i\phi}\psi_{e,o}$.

To understand the local conservation of the vector current in space-time,
it is useful to consider the two-dimensional framework of the 8-vertex
model and its relation to the Hamiltonian formalism. Each term $H_j$
in the Hamiltonian comes from the expansion of a local vertex connecting
sites $j$ and $j+1$. Denoting the fermion states on these two sites by
+ (occupied) and - (unoccupied), and using (4) and (5), we can write
(for $\Delta=0, k\approx 1$)
\begin{eqnarray}
H_j|++\rangle & = & \epsilon|--\rangle,\,\,\,
H_j|--\rangle  =  \epsilon|++\rangle \\
H_j|+-\rangle & = & |-+\rangle,\,\,\,
H_j|-+\rangle  =  |+-\rangle
\end{eqnarray}
where $\epsilon=(1-k)/2$ is proportional to the fermion mass.
If $j$ is even, the prescription (\ref{eq:evenodd}) for the Dirac field
corresponds to taking a new basis of states denoted by charge
numbers associated with densities $\psi_1^*\psi_1$ and $\psi_2^*\psi_2$,
\begin{eqnarray}
|00\rangle & = & \left(|+-\rangle - |-+\rangle\right)/\sqrt{2} \\
|11\rangle & = &\left(|+-\rangle + |-+\rangle\right)/\sqrt{2} \\
|01\rangle & = &|--\rangle, \,\,\, |10\rangle = |++\rangle
\end{eqnarray}
for which we have
\begin{eqnarray}
H_j|00\rangle & = &-|00\rangle,\,\,\,
H_j|01\rangle  = \epsilon|10\rangle \\
H_j|10\rangle & = &\epsilon|01\rangle,\,\,\,
H_j|11\rangle  = |11\rangle
\end{eqnarray}
In this second basis, the vector charge is the sum of the two occupation
numbers, and is thus  conserved through a vertex.

This work was supported in part by the U.S. Department of Energy under grant
No. DE-AS05-89ER40518.

%%%%%%%%%%%%%%%%%%%%%%%%%%%%%%%%%%%%%%%%%%%%%%%%%%%%%%%%%%%%%%%%%%%%%%%%

\end{document}